\documentclass[
preprint,
amsmath,
amssymb,
aps,
showkeys
]{revtex4-2}
\usepackage[compat=1.1.0]{tikz-feynman}
\tikzfeynmanset{warn luatex=false}
\usepackage{graphicx}% Include figure files
\usepackage{dcolumn}% Align table columns on decimal point
\usepackage{booktabs}
\usepackage{amsmath}
\usepackage{braket}
\usepackage{dsfont}

\newcommand{\ud}{\text{d}}

\begin{document}

%\preprint{APS/123-QED}

\title{Collisional Approach for Open Neutrino Systems}

%\author{A. de Gouv\^ea$^{1}$} \email{degouvea@northwestern.edu}

\author{W.L. Ribeiro} \email{wlribeiro@ufabc.edu.br}
\author{C.A. Moura} \email{celio.moura@ufabc.edu.br}

%\affiliation{$^{1}$Northwestern University, Department of Physics \& Astronomy, Evanston, IL 60208, USA}
\affiliation{Universidade Federal do ABC (UFABC), Santo Andr\'e, SP 09210-580, Brazil}

\date{\today}% It is always \today, today,

\begin{abstract}
We develop a collisional framework for neutrino propagation within open quantum systems, termed the \emph{Collisional Approach for Open Neutrino Systems} (CAONS). A Born-Markov equation is derived, linking decoherence, dissipation, decay rates, and scattering cross sections. Perturbation theory is not required and the resultant master equation is applied to visible and invisible neutrino decays and propagation through a stationary medium. Comparing with previous studies on neutrino decoherence, we show that current bounds on decoherence parameters can significantly tighten constraints on neutrino couplings with dark-matter. Lastly, we establish connections between CAONS and non-Hermitian Hamiltonian approaches.
\end{abstract}

\keywords{Neutrino Propagation, Open Quantum System, Decoherence, Born-Markov Equation, Non-Hermitian Quantum Mechanics}

\maketitle

\section{\label{sec:Introduction} Introduction}

The neutrino as an Open Quantum System (OQS) offers a novel approach to explore physics beyond the Standard Model (SM)~\cite{sun, lisi2000, hooper2005}. A key motivation for applying the OQS framework in neutrino physics is its ability to transform pure neutrino states into mixed states, i.e., from a single vector state into a statistical mixture. This loss of purity entails entropy production and irreversibility, unlike standard neutrino oscillation models, where purity is preserved.

The first application of the OQS framework to neutrinos dates back to the late 1990s, providing an alternative explanation for the atmospheric muon neutrino deficit~\cite{grossman}. That same year, Sun and Zhou~\cite{sun} explored gravity effects and black hole evaporation as motivations for employing OQS in neutrino physics.

The time evolution of systems weakly coupled to a \emph{memoryless} environment — where interactions are independent of previous events — is described by the Born-Markov master equation. When diagonalized, it yields the Gorini-Kossakowski-Sudarshan-Lindblad equation (Lindblad equation)~\cite{breuer, schlosshauer}. In 2000, Lisi \emph{et al.}~\cite{lisi2000} applied the Lindblad equation to neutrino phenomenology, proposing that the Super-Kamiokande experiment could detect decoherence effects caused by unknown physics, such as quantum gravity. Lacking a physical model for a microscopic derivation, they introduced \emph{ad hoc} decoherence parameters proportional to powers of the neutrino energy, assuming energy conservation and no dissipation. 

While decoherence can reveal new physics, it may also affect measurements of standard oscillation parameters, including the CP-violation phase and the mixing angle $\theta_{23}$~\cite{coelho2017, carpio2018}. Thus, the OQS approach is crucial not only for probing new interactions but also for correctly interpreting neutrino oscillation data.
In neutrino physics, an OQS framework based on scattering theory is especially appealing, as it aligns with quantum field theory—the most advanced framework for particle interactions.

Many studies link neutrino decoherence to quantum gravity~\cite{gago2002, fogli2003, hooper2005, barenboim2005, morgan2006, fogli2007, mavromatos2008, sakharov2009exploration, capolupo2019, buoninfante2020, hellmann2021}, using various experiments to constrain the decoherence parameters~\cite{carpio2018, sakharov2009exploration, fogli2003, gomes2017, coloma2018}. However, a microscopic derivation requires an interaction Hamiltonian and the environmental state, which together determine the decoherence and dissipation parameters, denoted generically as $\gamma$~\cite{breuer, schlosshauer}. Understanding neutrino propagation can also reveal properties of the environments they traverse, potentially altering interpretations of high-energy neutrino sources.

Different techniques exist to derive master equations. For systems with finite Hilbert spaces, the eigenoperator method is appropriate~\cite{breuer}. For scattering processes, however, the collisional method (or repeated interaction scheme) is often more suitable~\cite{hornberger, hornberger2}. We focus on two relevant approaches:  
\emph{i}) The first approach considers the complete evolution for a short time interval, $\Delta t$, expanding the time evolution operator and tracing over the environment~\cite{ciccarello}. However, this method introduces an undesirable dependence of $\gamma$ on the arbitrary time interval $\Delta t$.  
\emph{ii}) The second approach, developed by Hornberger and Sipe~\cite{hornberger}, applies scattering theory to model the evolution of Brownian particles interacting with their environment. They derived a cross section-dependent equation but introduced the unitary part by hand, an unexpected feature since free evolution should arise naturally from the full equation.

In this work, we develop a collisional approach to derive a Born-Markov master equation for neutrinos, termed the \emph{Collisional Approach for Open Neutrino Systems} (CAONS). This method overcomes the challenges of previous approaches: i) the $\Delta t$ dependence of the $\gamma$ parameters and ii) the arbitrary inclusion of the unitary part. Additionally, unlike most of the existing methods, CAONS is not perturbative, in the sense that the master equation does not come from perturbation theory, and provides a clear interpretation of $\gamma$ in terms of scattering cross sections and decay rates.  
Section~\ref{sec:Theory} presents the derivation of the CAONS framework. In Section~\ref{applications}, we apply it to neutrino decay and scattering. Section~\ref{implications} discusses the implications of our results, followed by final remarks in Section~\ref{conclusions}.

 \section{Collisional Approach for Open Neutrino Systems (CAONS)}\label{sec:Theory}

In this section, we develop the CAONS framework to study decoherence, particularly in neutrino systems. Although motivated by neutrino propagation, this method is general and can be applied to any particle system subject to decoherence or dissipation.  

Let $\rho_{tot}(0)$ denote the initial state of the composite neutrino-environment system in the Schrödinger picture, with $\nu$ representing the neutrino. Assuming the two systems are initially uncorrelated, we write  
\begin{equation}
   \rho_{tot}(0) = \rho(0) \otimes \rho^E(0)\,, 
\end{equation}
where $\rho(0)$ and $\rho^E(0)$ are the initial density operators of the neutrino and environment, respectively. Given the weak interaction between neutrinos and the environment, the macroscopic state of the environment remains unchanged, and the composite system can be approximated by a product state throughout the evolution—an assumption known as the Born approximation~\cite{schlosshauer}. We focus on free particles and vacuum environments, so $\rho^E$ commutes with the  free Hamiltonian. 

After evolving for a time interval $\Delta t$, the state becomes  
\begin{equation}
    \rho_{tot}(\Delta t) = U(\Delta t) \rho_{tot}(0) U^\dagger(\Delta t)\,,
    \label{eq:nu-env-schroed}
\end{equation}
where \( U(\Delta t) \) is the time-evolution operator. In the interaction picture, Eq.~(\ref{eq:nu-env-schroed}) transforms into  
\begin{equation}
    \rho_{tot}^I(\Delta t) = S_+(\Delta t) \rho_{tot}^I(0) S_+^\dagger(\Delta t)\,,
\end{equation}
where \( S_+ \) is the scattering matrix for non-negative times, defined as  
\begin{equation}
S_+(\Delta t) = \mathcal{T} \exp\left(-i \int_0^{\Delta t} \mathcal{H}_{int}(x) \, \mathrm{d}^4x\right),
\end{equation}
with \(\mathcal{T}\) denoting the time-ordering operator and \(\mathcal{H}_{int}(x)\) the interaction Hamiltonian coupling the system and environment.  

Expanding \( S_+ = \mathds{1} + i T_+ \), where \( T_+ \) is the non-negative time transfer matrix, we obtain  
\begin{equation} \label{Collision1}
   \Delta \rho_{tot}^I = i T_+ \rho_{tot}^I(0) - i \rho_{tot}^I(0) T_+^\dagger + T_+ \rho_{tot}^I(0) T_+^\dagger\,.
\end{equation}

It is important to note that we choose $\Delta t$ in Eqs.~\eqref{eq:nu-env-schroed}–\eqref{Collision1} to be small compared to the system's evolution timescale but large compared to the interaction time. This ensures that the upper limit of the time integral in $S_+$, and consequently in $T_+$, can extend to $\Delta t \to \infty$, introducing the Markov approximation. This approximation enables us to compute the matrix elements of $T_+$ similarly to the standard transfer matrix \( T \). Specifically, instead of  
\begin{equation}
    \braket{f|T|i} = (2\pi)^4 \delta^4(p_i^\mu - p_f^\mu) \, \mathcal{M},
\end{equation}
we obtain  
\begin{align}
    \braket{f|T_+|i} = (2\pi)^3 \delta^3(\mathbf{p_i} - \mathbf{p_f}) 
    \Bigg[\pi \delta(E_i - E_f) - i\, \text{P.V.}\left(\frac{1}{E_i - E_f}\right)\Bigg] \mathcal{M},
\end{align}
where P.V. denotes the Cauchy principal value.

Since $S_+$ is unitary, i.e., \( S_+^\dagger S_+ = \mathds{1} \), it follows that  
\begin{align}
    i T_+^\dagger &= i T_+ + T_+^\dagger T_+, \\
    i T_+ &= i T_+^\dagger + T_+^\dagger T_+.
\end{align}

Applying these results to Eq.~(\ref{Collision1}) and dividing by $\Delta t$, we obtain
\begin{equation}
    \frac{\Delta \rho_{tot}^I}{\Delta t} = \frac{1}{\Delta t} \left( -i \left[\frac{T_+ + T_+^\dagger}{2}, \rho_{tot}(0)\right] + T_+ \rho_{tot}(0) T_+^\dagger - \frac{1}{2}\{T_+^\dagger T_+, \rho_{tot}(0)\}\right).
\end{equation}
Taking the limit $\Delta t \to 0$ and reverting to the Schrödinger picture, then performing a partial trace over the environment, we find
\begin{equation}\label{eq:drhodt}
    \frac{\ud \rho}{\ud t} = -i[H_0 + V, \rho] + \mathcal{D}(\rho),
\end{equation}
where $H_0$ is the free Hamiltonian of the neutrino, and
\begin{equation}\label{Potential}
    V = \lim_{\Delta t \to 0} U_0 \left(\frac{\langle T_+ + T_+^\dagger \rangle_E}{2 \Delta t}\right) U_0^\dagger,
\end{equation}
with $U_0$ being the time-evolution operator for the free neutrino and $\langle \cdot \rangle_E$ denoting the average over the environment states. In Eq.~\eqref{eq:drhodt},
\begin{equation}\label{fakedissipator}
    \mathcal{D}(\rho) = \lim_{\Delta t \to 0} \frac{1}{\Delta t} U_0 \text{tr}_E \left(T_+ \rho_{tot} T_+^\dagger - \frac{1}{2}\{T_+^\dagger T_+, \rho_{tot}\}\right) U_0^\dagger.
\end{equation}
Although $\mathcal{D}(\rho)$ is defined as in Eq.~(\ref{fakedissipator}), it requires corrections to ensure unitarity. Specifically, the imaginary part must satisfy $\text{Im} \, \mathcal{D}(\rho) = -[H_{LS}, \rho]$, where $H_{LS}$ is the Lamb-shift Hamiltonian. Thus, we redefine $\mathcal{D}(\rho) \rightarrow \text{Re} \, \mathcal{D}(\rho)$ and include $H_{LS}$ in the unitary part of the evolution.

%%%%%%%%%%%%%%%%%%%%%%%%%%%%%%%%%%%%%%%%%
There are three key points regarding the master equation:  
\emph{i}) The initial time $t = 0$ is arbitrary, so the equation must hold for any $t \geq 0$.  
\emph{ii}) Although $\Delta t$ in $V$ and $\mathcal{D}(\rho)$ appears divergent, it is regularized in $V$ and governs the emergence of probability rates in the dissipator.  
\emph{iii}) Unlike standard Born-Markov equations, our result is not perturbative by default, though perturbation theory can be applied to the dissipation and decoherence parameters derived later.  

To express the master equation using operators acting solely on neutrino states, we solve the trace over the environment. Let $\{\ket{A}\}$ be a complete set of orthonormal states for the environment, with labels $A$ and $B$ for environment states, and $I, J, N,$ and $M$ for neutrino states. These labels fully specify the system; for instance, $\ket{I} \equiv \ket{i,\mathbf{p}}$ denotes a neutrino with momentum $\mathbf{p}$ and state $i$, whether in the flavor or mass basis. This notation applies to states with any number of particles.  

Assuming a diagonal density operator for the environment:  
\begin{equation}\label{envidensity}
    \rho^E(t) = \sum_{A} \rho_{AA}^E(t) \ket{A} \bra{A},
\end{equation}
we solve the partial trace and apply the rotating wave approximation~\cite{breuer,schlosshauer} to eliminate fast oscillatory terms. The resulting dissipator is:  
\begin{equation}\label{dissipator}
    \mathcal{D}(\rho) = \sum_{N,M} \gamma_{NM}(t) \left( K_{NM} \rho K_{NM}^\dagger - \frac{1}{2} \{ K_{NM}^\dagger K_{NM}, \rho \} \right),
\end{equation}
where the operators and parameters are defined as:  
\begin{equation}
    K_{NM} \ket{I} = \delta_{IN} \ket{M},
\end{equation}
\begin{equation} \label{decoparameters}
    \gamma_{NM}(t) = \sum_{A,B} \rho_{AA}^E(t) \frac{P(N,A \to M,B)}{\Delta t}.
\end{equation}
The $\gamma_{NM}(t)$ parameters are probability rates averaged over the environment, with $P(N,A \to M,B)$ representing the transition probabilities.

%%%%%%%%%%%%%%%%%%%%%%%%%%%%%%
We emphasize that $\gamma$ parameters can be interpreted in terms of decay and scattering rates due to their connection with transition probabilities. CAONS fully determines the master equation structure, requiring only the probability rates to compute the neutrino time evolution within this framework.

Next, we briefly discuss $H_{LS}$ and $V$. The Lamb-shift Hamiltonian, $H_{LS}$, introduces a small correction to the system’s energy levels~\cite{breuer}. It commutes with the free Hamiltonian and has no significant influence on the evolution \emph{a priori}. Thus, we absorb $H_{LS}$ into the free Hamiltonian in this study. 

Focusing on $V$, we expand $T_+ (T_+^\dagger)$ to first order using Eq.~(\ref{Potential}):
\begin{align}
    V &\approx \lim_{\Delta t \to 0} \sum_A \frac{\rho_{AA}(t)}{\Delta t} \bra{A} \int \ud^4x \, \mathcal{H}(x) \ket{A} \nonumber \\
    & = \sum_A \rho_{AA}(t) \bra{A} \int \ud^3x \, \mathcal{H}(x) \ket{A} = \braket{H_{int}}.
\end{align}
Thus, $V$ corresponds to the matter potential in neutrino propagation, similar to the case of solar neutrinos~\cite{giunti}.

%%%%%%%%%%%%%%%%%
\subsection{Evolution of a Subspace}
We now compute the time evolution of each element of $\rho$, with $\rho_{IJ} \equiv \braket{I|\rho|J}$:
\begin{align}
    \dot{\rho}_{IJ} &= \sum_N -i \left( H_{IN} \rho_{NJ} - H_{NJ} \rho_{IN} \right) + \delta_{IJ} \gamma_{NI} \rho_{NN} \nonumber \\
    &\quad - \frac{1}{2} (\gamma_{IN} + \gamma_{JN}) \rho_{IJ}.
\end{align}
The master equation yields an infinite set of coupled differential equations due to the combined effects of mass differences and momentum variations, unlike standard treatments focused solely on mass dynamics.

To simplify, we decompose the evolution into subspaces, aligning our formalism with standard approaches. Consider neutrinos with momentum $\mathbf{p}$ in the subspace $\mathcal{P} = \{\ket{n, \mathbf{p}}\}$, where $n$ denotes either mass or flavor states. Let $\mathds{P} = \sum_n \ket{n, \mathbf{p}} \bra{n, \mathbf{p}}$ be the projection operator for $\mathcal{P}$, and $\mathds{Q} = \mathds{1} - \mathds{P}$ for the complementary subspace.

We focus on scenarios where neutrinos either decay (in)visibly in vacuum or scatter on resting particles. In these cases, neutrinos may leave $\mathcal{P}$ but cannot return, with the approximation that environmental particles have negligible momentum.

Define the following components:
\begin{align} \label{defsubspace}
    \rho^{PP} &= \mathds{P} \rho \mathds{P}, & H^P &= \mathds{P} H \mathds{P}, \nonumber \\
    \rho^{QQ} &= \mathds{Q} \rho \mathds{Q}, & H^Q &= \mathds{Q} H \mathds{Q}, \nonumber \\
    \rho^{PQ} &= \mathds{P} \rho \mathds{Q}, & V^{PQ} &= \mathds{P} H \mathds{Q}, \nonumber \\
    \rho^{QP} &= \mathds{Q} \rho \mathds{P}, & V^{QP} &= \mathds{Q} H \mathds{P}.
\end{align}
Using these definitions, the density matrix takes the form:
\begin{equation}
    \rho = 
    \begin{pmatrix}
        \rho^{PP} & \rho^{PQ} \\
        \rho^{QP} & \rho^{QQ}
    \end{pmatrix}.
\end{equation}

The differential equations for each block are:
\begin{widetext}
\begin{align}
    \dot{\rho}^{PP} &= -i \left([H^P, \rho^{PP}] + V^{PQ} \rho^{QP} - \rho^{PQ} V^{QP} \right) 
    - \frac{1}{2} \sum_{N,M} \gamma_{NM}(t) \{ K_{NM}^{\dagger PP} K_{NM}^{PP}, \rho^{PP} \}, \\
    \dot{\rho}^{QQ} &= -i \left([H^Q, \rho^{QQ}] + V^{QP} \rho^{PQ} - \rho^{QP} V^{PQ} \right) 
    + \sum_{N,M} \gamma_{NM}(t) K_{NM}^{QP} \rho^{PP} K_{NM}^{\dagger PQ}, \\
    \dot{\rho}^{QP} &= -i \left(H^Q \rho^{QP} - \rho^{QP} H^P + V^{QP} \rho^{PP} - \rho^{QQ} V^{QP} \right) 
    - \frac{1}{2} \sum_{N,M} \gamma_{NM}(t) \rho^{QP} K_{NM}^{\dagger PQ} K_{NM}^{QP}.
\end{align}
\end{widetext}
The equation for $\rho^{PQ}$ is the Hermitian conjugate of $\rho^{QP}$. 

Next, we simplify $V^{PQ} = \mathds{P}(H_0 + V)\mathds{Q}$. With $H_0 = \sum_{n, \mathbf{k}} E_n(\mathbf{k}) \ket{n, \mathbf{k}} \bra{n, \mathbf{k}}$, we find:
\begin{align}
    \mathds{P} H_0 \mathds{Q} &= \sum_{n,l,m} \sum_{\mathbf{k}, \mathbf{q}} E_n(\mathbf{k}) \ket{l, \mathbf{p}} 
    \braket{l, \mathbf{p} | n, \mathbf{k}} \nonumber \\
    &\quad \times \braket{n, \mathbf{k} | m, \mathbf{q}} \bra{m, \mathbf{q}} \propto \delta_{\mathbf{p}, \mathbf{q}} = 0.
\end{align}
Similarly:
\begin{align}
    \mathds{P} V \mathds{Q} &= \lim_{\Delta t \to 0} \sum_{n,m,A} \sum_{\mathbf{q}} \frac{e^{-i \Delta t (E_n(\mathbf{p}) - E_m(\mathbf{q}))}}{\Delta t} \nonumber \\
    &\quad \times \ket{n, \mathbf{p}} \bra{m, \mathbf{q}} \mathcal{A}((m, \mathbf{q}), A \to (n, \mathbf{p}), A) \nonumber \\
    &\quad + \text{c.c.} = 0,
\end{align}
since $\mathbf{p}$ and $\mathbf{q}$ belong to orthogonal subspaces. Thus, $V^{PQ} = \mathds{P} H \mathds{Q} = 0$. The master equation for each block simplifies to:
\begin{align}
    \dot{\rho}^{PP} &= -i [H^P, \rho^{PP}] - \frac{1}{2} \sum_{N,M} \gamma_{NM}(t) \{ K_{NM}^{\dagger PP} K_{NM}^{PP}, \rho^{PP} \}, \label{PPequation} \\
    \dot{\rho}^{QQ} &= -i [H^Q, \rho^{QQ}] + \sum_{N,M} \gamma_{NM}(t) K_{NM}^{QP} \rho^{PP} K_{NM}^{\dagger PQ}, \label{rhoqqevo} \\
    \dot{\rho}^{QP} &= -i (H^Q \rho^{QP} - \rho^{QP} H^P) 
    - \frac{1}{2} \sum_{N,M} \gamma_{NM}(t) \rho^{QP} K_{NM}^{\dagger PQ} K_{NM}^{QP}.
\end{align}

If the momentum superposition is negligible, we take $\rho_{ij}(\mathbf{q}, \mathbf{p}; 0) = 0$, giving:
\begin{equation}
    \rho = 
    \begin{pmatrix}
        \rho^{PP} & 0 \\
        0 & \rho^{QQ}
    \end{pmatrix}.
\end{equation}
This result allows us to disregard $\rho^{QP}$ in further analysis.
 
%%%%%%%%%%%%%%%%%%%%%%%%%%%%%%%%%%%%%%%%%%
%%%%%%%%%%%%%%%%%%
\section{Basic Applications}\label{applications}

This section presents basic applications of the collisional approach, focusing on three examples:  
\emph{i) Neutrino invisible decay}: The simplest case, where the neutrino decays to its vacuum state.  
\emph{ii) Neutrino visible decay}: The decay results in less energetic neutrinos due to energy-momentum conservation, with final states tending to be less massive.  
\emph{iii) Neutrino scattering on particles at rest}: Similar to visible decay, with the neutrino interacting with stationary particles.  

For simplicity, we consider a two-family neutrino scenario. The flavor eigenstates, $\ket{\alpha,\mathbf{k}}$ and $\ket{\beta, \mathbf{k}}$, are related to the mass eigenstates $\ket{1,\mathbf{k}}$ and $\ket{2,\mathbf{k}}$ by:
\begin{equation}
    \begin{pmatrix}
        \ket{\alpha,\mathbf{k}} \\ \ket{\beta,\mathbf{k}}
    \end{pmatrix}
    = \begin{pmatrix}
        \cos(\theta) & \sin(\theta) \\
       -\sin(\theta) & \cos(\theta)
    \end{pmatrix}
    \begin{pmatrix}
        \ket{1,\mathbf{k}} \\ \ket{2,\mathbf{k}}
    \end{pmatrix}.
\end{equation}

The transition probabilities are computed using:
\begin{equation}\label{probformula}
P_{\nu_{\alpha,\mathbf{p}} \to \nu_{\beta,\mathbf{k}}}(t) = \text{tr}\big(\ket{\beta,\mathbf{k}}\bra{\beta,\mathbf{k}} \rho^{(\alpha,\mathbf{p})}(t)\big),
\end{equation}
where $\rho^{(\alpha,\mathbf{p})}(t)$ is the density matrix of a neutrino created in state $\ket{\alpha,\mathbf{p}}$.

%%%%%%%%%%%%%%
\subsection{Neutrino Invisible Decay}

Consider a neutrino $\nu_\gamma$ that decays invisibly. Its interaction Lagrangian is given by~\cite{pasquini2016,choubey2021exploring,gelmini1984fast}:
\begin{equation}
    \mathcal{L}(x) = g_\gamma \bar{\nu}_\gamma(x) E(x) + \text{h.c.},
\end{equation}
where $E(x)$ includes fields related to the decay. If $\nu_\gamma$ is a linear combination of mass eigenstates:
\begin{equation}
    \nu_\gamma(x) = \sum_i \Tilde{U}_{\gamma i} \nu_i(x),
\end{equation}
the Lagrangian becomes:
\begin{equation}
    \mathcal{L}(x) = \sum_i g_i \bar{\nu}_i(x) E(x) + \text{h.c.}
\end{equation}

\begin{figure}
    \centering
    \begin{tikzpicture}
        \begin{feynman}
            \vertex (a) {\(\nu_i\)};
            \vertex [right=of a] (b);
            \vertex [above right=of b] (c) {\(\phi\)};
            \vertex [below right=of b] (d) {\(F\)};
            \diagram* {
                (a) -- [fermion] (b),
                (c) -- [scalar] (b),
                (b) -- [fermion] (d),
            };
        \end{feynman}
    \end{tikzpicture}
    \caption{Feynman diagram for neutrino invisible decay.}
    \label{invisible}
\end{figure}
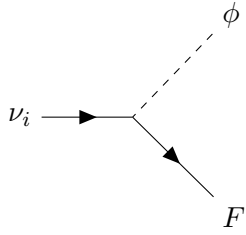

We assume a simple model with $E(x) = F(x)\phi(x)$, where $F(x)$ is a fermionic field and $\phi(x)$ is a real scalar. Fig.~\ref{invisible} shows the Feynman diagram for the invisible decay. In a vacuum environment, $\rho_{AA} = 1$ if $A$ is the vacuum state, and the final neutrino state is the vacuum ($M = 0$ in Eq.~\ref{decoparameters}). The total decay rate $\gamma_{N0}$ simplifies to:
\begin{equation}
    \gamma_{N0} = \Gamma(N),
\end{equation}
where $\Gamma(N)$ is the total decay rate for a neutrino in state $N$.

For a neutrino created in state $\ket{\alpha,\mathbf{p}}$, the initial density matrix is:
\begin{equation}
    \rho^{(\alpha,\mathbf{p})}(0) = \begin{pmatrix}
        \cos^2(\theta) & \sin(\theta)\cos(\theta) \\
        \sin(\theta)\cos(\theta) & \sin^2(\theta)
    \end{pmatrix}.
\end{equation}
The evolution of $\rho_{ij} = \braket{i,\mathbf{p}|\rho|j,\mathbf{p}}$, provided by Eq.(\ref{PPequation}) is:
\begin{equation}
    \dot{\rho_{ij}}(t) = -\left[i\Delta_{ij}(\mathbf{p}) + \frac{\Gamma_i(\mathbf{p}) + \Gamma_j(\mathbf{p})}{2}\right] \rho_{ij}(t),
\end{equation}
where $\Delta_{ij}(\mathbf{p}) = \frac{\Delta m^2_{ij}}{2|\mathbf{p}|}$. Thus, the solution is:
\begin{equation}
    \rho_{ij}(t) = \rho_{ij}(0) \exp\left(-\left[i\Delta_{ij}(\mathbf{p}) + \frac{\Gamma_i(\mathbf{p}) + \Gamma_j(\mathbf{p})}{2}\right]t\right).
\end{equation}

The evolved density matrix is:
\begin{widetext}
\begin{equation}\label{rhoinvdecay}
    \rho^{(\alpha,\mathbf{p})}(t) = \begin{pmatrix}
        \cos^2(\theta)e^{-\Gamma_1 t} & \frac{\sin(2\theta)}{2} e^{-(-i\Delta_{21} + (\Gamma_1 + \Gamma_2)/2)t} \\
        \frac{\sin(2\theta)}{2} e^{-(i\Delta_{21} + (\Gamma_1 + \Gamma_2)/2)t} & \sin^2(\theta)e^{-\Gamma_2 t}
    \end{pmatrix}.
\end{equation}
\end{widetext}

The survival probability is:
\begin{align}
    P_{\nu_{\alpha}(\mathbf{p})\to \nu_{\alpha}(\mathbf{p})}(t) &= \cos^4(\theta) e^{-\Gamma_1 t} + \sin^4(\theta) e^{-\Gamma_2 t} \nonumber \\
    &\quad + \frac{1}{2} \sin^2(2\theta) \cos(\Delta_{21} t) e^{-(\Gamma_1 + \Gamma_2)/2 t}.
\end{align}

\begin{figure}
    \centering
    \includegraphics[scale=0.62]{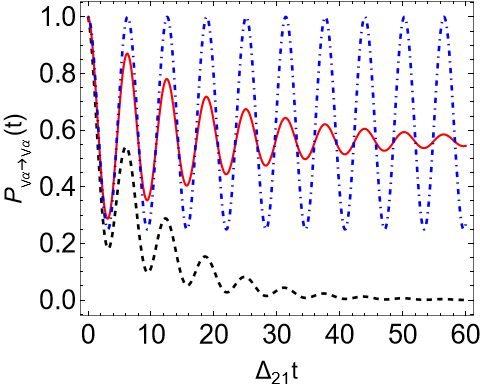}
    \caption{(Color Online) Survival probability for 2-family neutrino oscillations as a function of $\Delta_{21}t$. Blue dot-dashed: standard case (no decay). Solid red: $\nu_1$ decays. Dashed black: both eigenstates decay. We use $\Gamma = 0.1 \Delta_{21}$ and $\theta = \pi/3$.}
    \label{fig:PSID}
\end{figure}

Fig.~\ref{fig:PSID} shows survival probabilities for different decay rates, including the standard case (no decay), $\Gamma_1 \approx \Gamma_2 = 0.1 \Delta_{21}$, and $\Gamma_1 = 0.1 \Delta_{21}$ with $\Gamma_2 = 0$. As expected, if both mass eigenstates decay, the survival probability tends to zero as $t \to \infty$. If only one eigenstate decays, the asymptotic probability equals the initial population of the other mass eigenstate.

\subsection{Neutrino Visible Decay}

We now consider the visible decay of neutrinos in vacuum. Unlike invisible decay, visible decay changes the proportion of mass eigenstates rather than making neutrinos disappear. Additionally, visible decay alters the neutrino’s momentum, requiring the computation of matrix elements in the $\mathcal{Q}$ subspace.

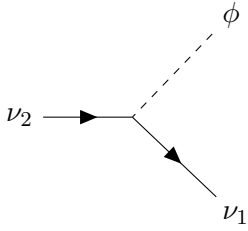
\begin{figure}
    \centering
    \begin{tikzpicture}
        \begin{feynman}
            \vertex (a) {\(\nu_2\)};
            \vertex [right=of a] (b);
            \vertex [above right=of b] (c) {\(\phi\)};
            \vertex [below right=of b] (d) {\(\nu_1\)};
            \diagram* {
                (a) -- [fermion] (b),
                (c) -- [scalar] (b),
                (b) -- [fermion] (d),
            };
        \end{feynman}
    \end{tikzpicture}
    \caption{Tree-level Feynman diagram for neutrino visible decay.}
    \label{visible}
\end{figure}

Assume that a neutrino $\nu_2$ decays into $\nu_1$ plus a real scalar. The Lagrangian density is:
\begin{equation}
    \mathcal{L}(x) = g_{12} \bar{\nu}_2(x) \nu_1(x) \phi(x) + \text{h.c.}
\end{equation}
The tree-level Feynman diagram is shown in Fig.~\ref{visible}. In this scenario, only $\nu_2$ decays, so we simplify notation by writing $\Gamma$ instead of $\Gamma_2$.

Following the same procedure as for invisible decay, we find the evolved density matrix:
\begin{equation}\label{rhovisibledecay}
\rho^{(\alpha,\mathbf{p})}(t) = \begin{pmatrix}
    \cos^2(\theta) & \frac{\sin(2\theta)}{2} e^{(i\Delta_{21} - \Gamma/2)t} \\
    \frac{\sin(2\theta)}{2} e^{-(i\Delta_{21} + \Gamma/2)t} & \sin^2(\theta) e^{-\Gamma t}
\end{pmatrix}.
\end{equation}

The survival probability is:
\begin{align}
    P_{\nu_{\alpha}(\mathbf{p}) \to \nu_{\alpha}(\mathbf{p})}(t) &= \cos^4(\theta) + \sin^4(\theta) e^{-\Gamma t} \nonumber \\
    &\quad + \frac{1}{2} \sin^2(2\theta) \cos(\Delta_{21} t) e^{-\Gamma t/2}.
\end{align}
The asymptotic value of  $P_{\nu_{\alpha}(\mathbf{p}) \to \nu_{\alpha}(\mathbf{p})}(t \to \infty) = \cos^4(\theta)$. 

\begin{figure}
    \centering
    \includegraphics[scale=0.62]{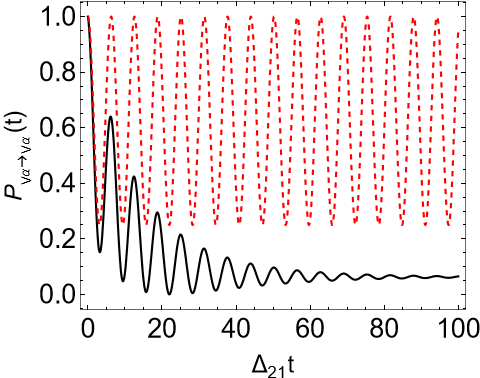}
    \caption{(Color Online) Survival probability as a function of $\Delta_{21}t$ for the neutrino visible decay. Red dashed: standard case. Solid black: neutrino visible decay with $\Gamma = 0.1 \Delta_{21}$ and $\theta = \pi/3$. The population $\rho_{11} (\mathbf{q},t)$ are not included.}
    \label{fig:PSVD}
\end{figure}

Fig.~\ref{fig:PSVD} shows the survival probabilities for this case, including the standard case (no decay) and visible decay with $\Gamma = 0.1 \Delta_{21}$ and $\theta = \frac{\pi}{3}$. As one can see, the probability converges to $\cos^4 (\frac{\pi}{3}) = 0.0625$. 

After computing the evolution of $\rho^{PP}$, we can compute the evolution for $\rho^{QQ}$. From Eq.~(\ref{rhoqqevo}), we have:
\begin{equation}
    \dot{\rho}_{ij}(\mathbf{q}, \mathbf{q'}, t) = \delta_{ij} \delta_{\mathbf{q}, \mathbf{q'}} \sum_n \gamma_{n,i}(\mathbf{p}, \mathbf{q}) \rho_{nn}(\mathbf{p}, \mathbf{p}, t).
\end{equation}
Since only $\nu_2$ decays to $\nu_1$, we find:
\begin{equation} \label{eq:rho11}
    \rho_{11}(\mathbf{q}, t) = \frac{\gamma_{21}(\mathbf{p}, \mathbf{q})}{\Gamma} \sin^2(\theta) (1 - e^{-\Gamma t}),
\end{equation}
representing the population of neutrinos with mass $m_1$ and momentum $\mathbf{q}$.

The fraction in Eq.~\eqref{eq:rho11} represents the portion of neutrinos initially in mass eigenstate 2 with momentum $\mathbf{p}$ that decayed into eigenstate 1 with momentum $\mathbf{q}$. In practice, this fraction would be zero due to phase-space considerations. However, experimental measurements have finite precision, so integration over momentum intervals defined by experimental bins is necessary.
%%%

\subsection{Neutrino Scattering} \label{secNeutrinoScattering}

\begin{figure}
    \centering
    \begin{tikzpicture}
        \begin{feynman}
            \vertex (a) at (0,0);
            \vertex [above left=of a] (b) {\(\nu _i \)};
            \vertex [above right=of a] (c) {\(\phi\)};
            \vertex [below right=of a] (d) {\(\nu _j \)};
            \vertex [below left=of a] (e) {\(\phi\)};
            \diagram* {
                (b) -- [fermion] (a),
                (c) -- [scalar] (a),
                (a) -- [fermion] (d),
                (e) -- [scalar] (a),
            };
        \end{feynman}
    \end{tikzpicture}
    \caption{Interaction causing decoherence and dissipation in neutrino evolution. $\phi$ is a scalar particle and $\nu_{i,j}$ are neutrino mass eigenstates.}
    \label{fig:nuscattering}
\end{figure}
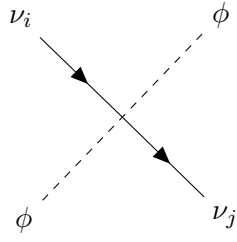

In this scenario, a neutrino propagates through a homogeneous environment composed of particles at rest. This case combines features from the previous examples: there is suppression of population terms, similar to Eq.~\eqref{rhoinvdecay}, but neutrinos remain in the final states. Consider the Lagrangian:
\begin{equation}
    \mathcal{L}(x) = \sum_{i,j} g_{ij} \bar{\nu}_i(x) \nu_j(x) \phi^2(x) + \text{h.c.},
\end{equation}
where $\phi(x)$ is a real scalar field. The interaction vertex is shown in Fig.~\ref{fig:nuscattering}.

The matter potential is:
\begin{equation}\label{matterpotential}
    V = \frac{n_\phi}{m_\phi |\mathbf{p}|} \begin{pmatrix}
        g_{11} m_1 & g_{12} \frac{m_1 + m_2}{2} \\
        g_{12}^* \frac{m_1 + m_2}{2} & g_{22} m_2
    \end{pmatrix},
\end{equation}
where $m_\phi$ is the scalar particle mass, and $n_\phi$ is its number density. We assume $g_{i \neq j} \ll g_{i = j}$, so the interacting eigenstates remain approximately the same as the free ones, but still allow mass-state transitions due to scattering.

The transition probability rates are:
\begin{equation}
    \frac{P(N,A \to M,B)}{\Delta t}(t) = n(t) |\mathbf{v}_{\text{rel}}| \sigma(N,A \to M,B),
\end{equation}
where $\sigma(N,A \to M,B)$ is the cross section, and $|\mathbf{v}_{\text{rel}}| \approx c$ is the relative velocity for ultra-relativistic neutrinos. Therefore, for the simple case where all environment particles are at rest, then 
\begin{equation} \label{gammascattering}
    \gamma_{NM} = n.\sigma (N \to M).
\end{equation}

The density matrix for the subspace $\mathcal{P}$ is similar to Eq.~\eqref{rhoinvdecay}, but with decay rates replaced by cross sections:
\begin{widetext}
\begin{equation}
    \rho^{(\alpha,\mathbf{p})}(t) \approx \begin{pmatrix}
        \cos^2(\theta) e^{-n \sigma_1(\mathbf{p}) t} & \frac{\sin(2\theta)}{2} e^{-(-i\Delta_{21} + n (\sigma_1(\mathbf{p}) + \sigma_2(\mathbf{p}))/2) t} \\
        \frac{\sin(2\theta)}{2} e^{-(i\Delta_{21} + n (\sigma_1(\mathbf{p}) + \sigma_2(\mathbf{p}))/2) t} & \sin^2(\theta) e^{-n \sigma_2(\mathbf{p}) t}
    \end{pmatrix}.
\end{equation}
\end{widetext}

Finally, the matrix elements for $\rho^{QQ}$ are:
\begin{equation}
    \rho_{jj}(\mathbf{q},t) = \sum_i \frac{\sigma_{ij}(\mathbf{p}, \mathbf{q})}{\sigma_{ij}(\mathbf{p})} \rho_{ii}(\mathbf{p}, 0) (1 - e^{-n \sigma_{i,\text{tot}}(\mathbf{p}) t}),
\end{equation}
which gives the population of neutrinos scattered into different mass eigenstates. This result is valid when $n \sigma c t \lesssim 1$; otherwise, multiple scatterings occur, making it difficult to separate subspaces. However, this provides a useful first approximation for the full master equation.
%%%%%%%%%%%%%%%%%%%%%%%%%%%%%%%

\section{Some Implications} \label{implications}

This section discusses indirect implications of our results, including estimates of weak interaction-induced decoherence, limits on neutrino-dark matter coupling constants, and the connection between CAONS and the non-Hermitian Hamiltonian approach.

As noted in Sec.~\ref{sec:Introduction}, early studies modeled decoherence parameters as proportional to some power of the neutrino energy, with bounds determined for different spectral indices. However, as shown in Eq.~\eqref{decoparameters}, $\gamma_{N,M}$ corresponds to environment-averaged probability rates. For simple environments, such as vacuum or particles at rest, $\gamma_{N,M}$ can be expressed in terms of decay rates and cross sections, which often scale with the neutrino energy.

For the analyses in Secs.~\ref{implicationWeak} and \ref{implicationDM}, we follow Ref.~\cite{coloma2018}, which offers a detailed study of Earth’s matter density profile. In their model, neutrino dynamics depend on three decoherence parameters expressed as $\gamma = \gamma_0 (E/E_0)^n$, where \(n\) is an integer in the range \([-2, 2]\).

\subsection{Weak-Interaction Induced Neutrino Decoherence} \label{implicationWeak}

As discussed in Sec.~\ref{secNeutrinoScattering}, for neutrinos scattering off particles at rest, the decoherence parameter $\gamma$ can be expressed as $n \sigma$. Using Earth’s density and the sub-TeV neutrino-nucleon cross section~\cite{formaggio2012}, we estimate $\gamma$ for atmospheric neutrinos traversing the Earth.

The Earth’s average density is $5.5 \, \text{g/cm}^3$, corresponding to a nucleon number density of \( n \sim 3.3 \times 10^{24} \, \text{cm}^{-3} \). The sub-TeV cross section is approximated by~\cite{formaggio2012}:
\begin{equation}
    \frac{\sigma_{\nu N}}{E} \approx 0.7 \times 10^{-38} \, \text{cm}^2 \, \text{GeV}^{-1}.
\end{equation}
Thus, using $\gamma = n.\sigma$, the decoherence parameter in natural units becomes:
\begin{equation}\label{eq:gammaweak}
    \gamma \approx 4.2 \times 10^{-28} \left(\frac{E}{E_0} \right) \, \text{GeV}, \quad E_0 = 1 \, \text{GeV}.
\end{equation}
This result aligns with Ref.~\cite{nieves2020}, indicating that most long-baseline experiments are sensitive only to $\gamma \sim 10^{-24}$ GeV, suggesting that hadronic matter decoherence may not significantly impact these experiments. However, Ref.~\cite{coloma2018} provides limits for $\gamma$ with $n = 1$ ranging from $3.5 \times 10^{-28}$ GeV to $3.3 \times 10^{-24}$ GeV, depending on the scenario. These values suggest that weak interaction-induced decoherence could be comparable to current bounds and should not be disregarded in searches for new physics.

\begin{figure}
    \centering
    \includegraphics[scale=0.55]{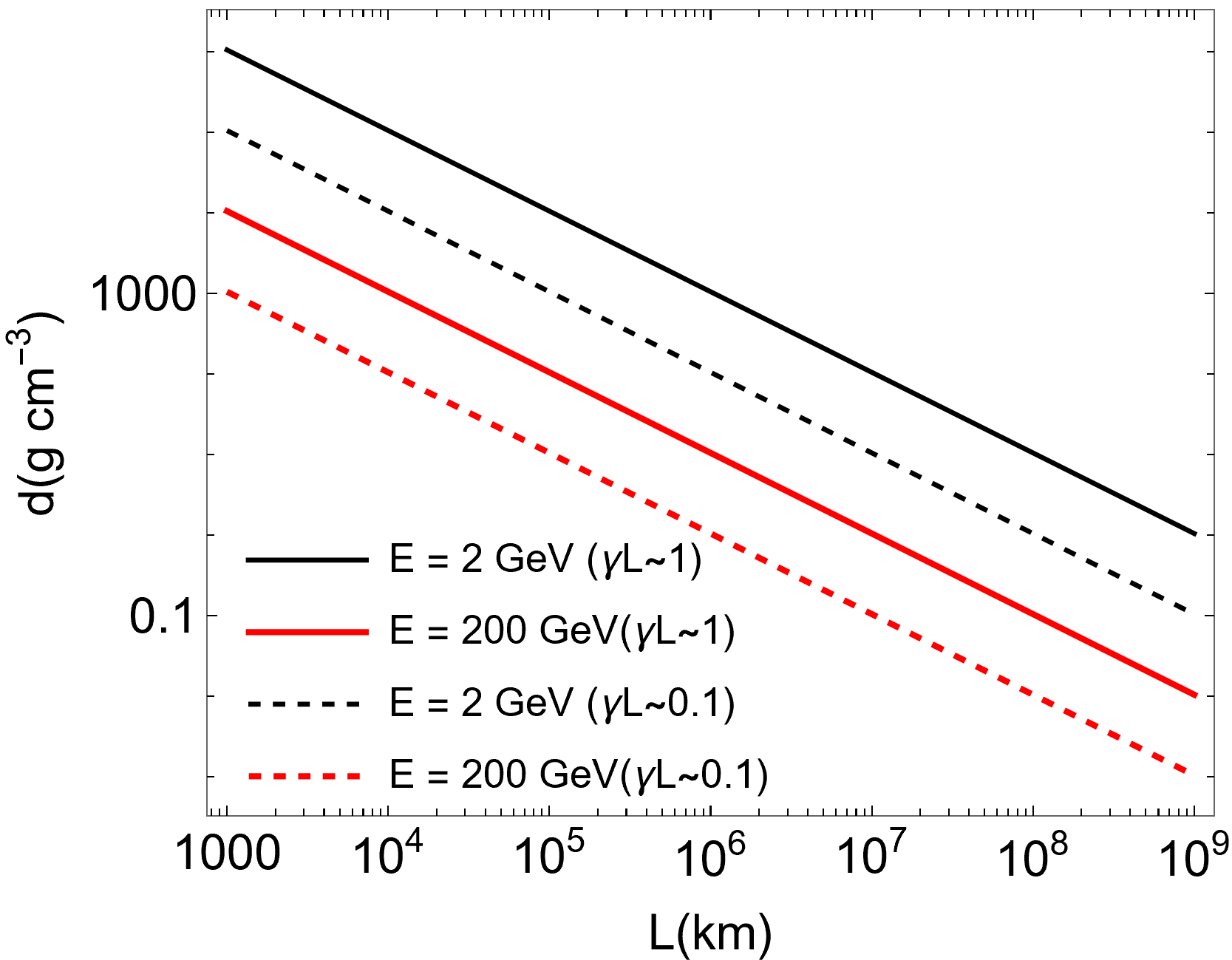}
    \caption{(Color online) Minimum matter density required for decoherence as a function of propagation distance. Solid lines: $\gamma L \sim 1$. Dashed lines: $\gamma L \sim 0.1$. Black: $E = 2$ GeV. Red: $E = 200$ GeV.}
    \label{fig:densityplot}
\end{figure}

Fig.~\ref{fig:densityplot} shows the relation between medium density and the neutrino baseline for different energies and values of $\gamma L$, where $\gamma$ is described by the Eq.(\ref{eq:gammaweak}). For $\mathcal{O}(1 \, \text{GeV})$ neutrinos, open dynamics require high-density environments or large objects. In contrast, $\mathcal{O}(100 \, \text{GeV})$ neutrinos experience decoherence even in stars like the Sun (radius $\sim 10^6$ km, average density $1.4 \, \text{g/cm}^3$~\cite{factbook}). Additionally, Ref.~\cite{coloma2018} suggests that for neutrinos crossing the Earth, $\gamma L \lesssim 10^{-2}$, indicating that both matter density and propagation length can be smaller than the values shown in Fig.~\ref{fig:densityplot}.

\subsection{Interactions with Scalar Dark Matter} \label{implicationDM}

Neutrino interactions with ultra-light scalar dark matter, or fuzzy dark matter (FDM), have been studied in~\cite{barranco2011,sen2023,huang2022}. In Ref.~\cite{barranco2011}, interactions are mediated by a neutral fermion of mass $M_I$, with FDM being either real (self-conjugate) or complex (non-self-conjugate). For real FDM and $s, u \ll M_I$ (Mandelstam variables), the cross section is:
\begin{equation}
    \sigma \approx \left(\frac{g_{\nu\phi}}{M_I}\right)^4 \frac{m_\nu^2}{16 \pi},
\end{equation}
where $g_{\nu\phi}$ is the neutrino-FDM coupling constant, and $m_\nu$ is the neutrino mass. For complex FDM:
\begin{equation}
    \sigma \approx \left(\frac{g_{\nu\phi}}{M_I}\right)^4 \frac{m_\phi E}{16 \pi},
\end{equation}
where $m_\phi$ is the FDM particle mass and $E$ is the neutrino energy.

Using the local dark matter energy density $\epsilon = 0.4 \, \text{GeV}/\text{cm}^3$~\cite{sofue2020} and the bounds on $\gamma$ from Ref.~\cite{coloma2018}, we derive upper limits on the coupling ratios. For complex FDM:
\begin{equation}
    \frac{g_{\nu\phi}}{M_I} < \left(\frac{16 \pi \gamma}{\epsilon E_0} \right)^{1/4}, \quad E_0 = 1 \, \text{GeV},
\end{equation}
and for real FDM:
\begin{equation}
    \frac{g_{\nu\phi}}{M_I} < \left(\frac{16 \pi \gamma m_\phi}{\epsilon m_\nu^2} \right)^{1/4}.
\end{equation}

\begin{figure}
    \centering
    \includegraphics[scale=0.62]{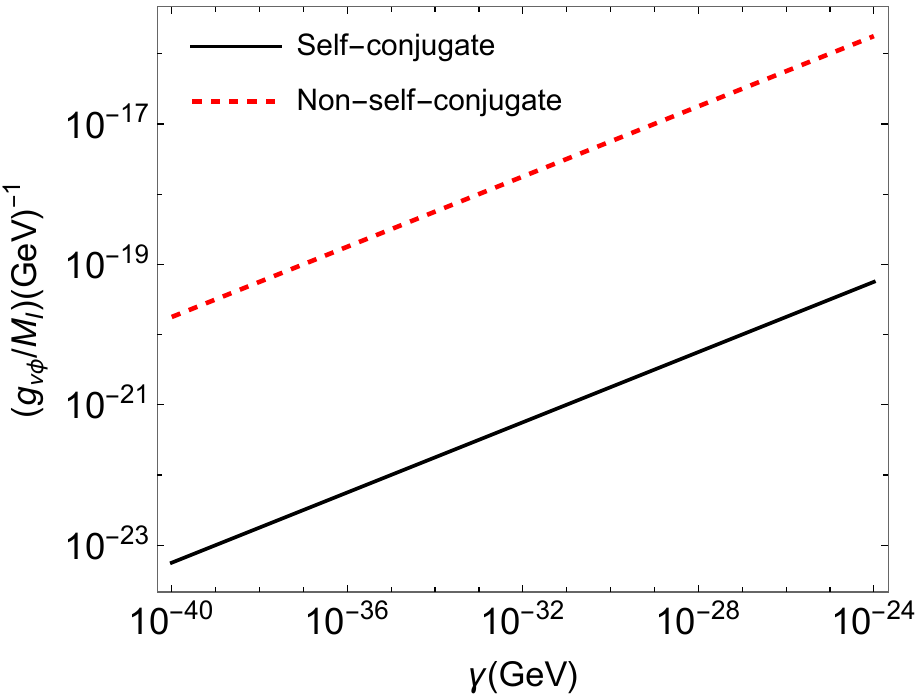}
    \caption{(Color online) $g_{\nu\phi}/M_I$ as a function of $\gamma$. The solid black line corresponds to real FDM, and the red dashed line to complex FDM. The shaded region shows allowed values for $g_{\nu\phi}/M_I$ based on current $\gamma$ limits.}
    \label{fig:g2}
\end{figure}

Using the results from Ref.~\cite{coloma2018}, Fig.~\ref{fig:g2} displays the constraints on \(g_{\nu\phi}/M_I\). While previous studies~\cite{barranco2011} set the ratio within \(10^{-2} - 10^{-1} \, \text{GeV}^{-1}\), our CAONS-based limits are significantly stricter, ranging from \(10^{-20}\) to \(10^{-16} \, \text{GeV}^{-1}\). This highlights the importance of linking decoherence parameters with cross sections to better understand neutrino interactions.

\subsection{Connection with the Non-Hermitian Hamiltonian} 

The open quantum system (OQS) formalism connects with the non-Hermitian Hamiltonian approach, where:
\begin{equation}\label{nonhermitiandef}
    H = (H_0 + V) - \frac{i}{2} \Gamma,
\end{equation}
with $\Gamma$ governing non-conservative dynamics, commonly used in decay and absorption studies~\cite{moiseyev2011}. 

The Lindblad equation is:
\begin{equation} \label{lindblad}
    \dot{\rho}(t) = -i[H, \rho] + \sum_n \gamma_n \left(L_n \rho L_n^\dagger - \frac{1}{2} \{L_n^\dagger L_n, \rho\}\right).
\end{equation}
Dropping the first term in the dissipator recovers the same dynamics as the non-Hermitian Hamiltonian:
\begin{equation} \label{nonhermitian}
    \Gamma \equiv \sum_n \gamma_n L_n^\dagger L_n.
\end{equation}

Thus, $\Gamma$ contains the same information about interactions and the environment as $\gamma_n$. Comparing Eq.~\eqref{nonhermitian} with Eq.~\eqref{PPequation}, we obtain:
\begin{equation} \label{gammaCAONS}
    \Gamma = \sum_{N,M} \gamma_{NM}(t) K_{NM}^\dagger K_{NM} = \sum_{N,M} \gamma_{NM}(t) K_{NN}.
\end{equation}

For diagonal $V$, the non-Hermitian and OQS approaches yield similar results. However, for non-diagonal $V$, the OQS method requires solving $\mathcal{O}(N^2)$ (likely coupled) equations, whereas the non-Hermitian approach only requires $N$ equations. Using Eqs.~\eqref{nonhermitian} and \eqref{gammaCAONS}, the Hamiltonian can be expressed as:
\begin{equation}
    H = \begin{pmatrix}
        H_1 & V \\
        V & H_2
    \end{pmatrix} - \frac{i}{2} \begin{pmatrix}
        \gamma_1 & 0 \\
        0 & \gamma_2
    \end{pmatrix}.
\end{equation}

After diagonalizing $(H_0 + V)$, we get:
\begin{equation}
    \Tilde{H} = \begin{pmatrix}
        \Tilde{H}_1 & 0 \\
        0 & \Tilde{H}_2
    \end{pmatrix} - \frac{i}{2} \begin{pmatrix}
        \Tilde{\gamma}_{11} & \Tilde{\gamma}_{12} \\
        \Tilde{\gamma}_{21} & \Tilde{\gamma}_{22}
    \end{pmatrix}.
\end{equation}

Thus, CAONS provides a tool for uncovering the microscopic meaning of $\Tilde{\gamma}_{ij}$, as it directly relates to transition probability rates.

%%%%%%%%%%%%%%%%%%
\section{Conclusions and Final Remarks}\label{conclusions}
Using quantum scattering theory, we derived a general Born-Markov master equation, linking the \(\gamma\) parameters to decay rates and cross sections. The resulting equation is not perturbative, setting it apart from most methods in the literature. A key advantage of the CAONS framework is its flexibility: \(\gamma\) can be computed from interaction models and environmental particle profiles or inferred from phenomenological transition rates, making it especially useful when kinematics play a central role. 

The CAONS framework enables precise determination of neutrino time evolution, decay rates, and energy dissipation through scattering processes. Additionally, the relation between \(\gamma\) and cross sections provides valuable insights into neutrino-environment interactions. One important finding is that decoherence caused by hadronic matter lies within established limits, highlighting the need to account for it when searching for non-standard neutrino interactions. If neutrino decoherence is observed, contributions from hadronic matter must be considered to accurately estimate the effect of non-standard interactions.  

We also applied the CAONS framework to constrain the coupling between neutrinos and ultra-light dark matter. Current limits on \(\gamma\) yield a much stricter constraint on \(g_{\nu \phi}/M_I\), with $g_{\nu \phi}/M_I \approx 10^{-16} \, (10^{-20}) \, \text{GeV}^{-1}$ for complex (real) scalar dark matter—tightening the bound by at least 15 orders of magnitude compared to previous studies. This demonstrates the framework’s power in refining our understanding of neutrino interactions.

Finally, we showed how CAONS connects to the non-Hermitian Hamiltonian method, useful for studying irreversible particle disappearance. This connection underscores CAONS’s potential to enhance the application of non-Hermitian approaches across various contexts.

\section{Acknowledgements}
We would like to thank A. de Gouv\^ea for his comments and careful reading of the paper. This study was financed in part by the Coordenação de Aperfeiçoamento de Pessoal de Nível Superior - Brasil (CAPES) - Finance Code 001.  

\bibliography{referencesquali.bib}

\end{document}